\documentclass[preprint,showpacs,preprintnumbers,amsmath,amssymb,12pt]{revtex4-1}
\usepackage{lipsum}
\usepackage{setspace}
\usepackage{bm}
\usepackage[colorlinks,linkcolor = black,anchorcolor = blue,urlcolor = blue,	citecolor = blue]{hyperref}
\usepackage{float}
\usepackage{graphicx}
\usepackage{makecell}
\usepackage{lineno}
\usepackage{siunitx}

\begin{document}

\title{Persistent Fermi Pockets and Robust Electron Pairing in Lightly Doped CuO$_2$ Planes of Cuprate Superconductors}

\author{Hao Chen$^{1,2,\dag}$, Jumin Shi$^{1,2,\dag}$, Yinghao Li$^{1,2,\dag}$, Xiangyu Luo$^{1,\dag}$, Yiwen Chen$^{1,2}$, Chaohui Yin$^{1}$, Yingjie Shu$^{1,2}$, Jiuxiang Zhang$^{1}$, Taimin Miao$^{1,2}$, Bo Liang$^{1,2}$, Wenpei Zhu$^{1,2}$, Neng Cai$^{1,2}$, Xiaolin Ren$^{1,2}$, Chengtian Lin$^{3}$, Shenjin Zhang$^{4}$, Zhimin Wang$^{4}$, Fengfeng Zhang$^{4}$, Feng Yang$^{4}$, Qinjun Peng$^{4}$, Zuyan Xu$^{4}$, Guodong Liu$^{1,2,5}$, Hanqing Mao$^{1,2,5}$, Xintong Li$^{1,2,5}$, Tao Xiang$^{1,2,5,6}$, Lin Zhao$^{1,2,5,*}$ and X. J. Zhou$^{1,2,5,*}$}

\affiliation{
   \\$^{1}$Beijing National Laboratory for Condensed Matter Physics, Institute of Physics, Chinese Academy of Sciences, Beijing 100190, China.
   \\$^{2}$University of Chinese Academy of Sciences, Beijing 100049, China.
   \\$^{3}$Max Planck Institute for Solid State Research, Heisenbergstrasse 1, D-70569 Stuttgart, Germany.
   \\$^{4}$Technical Institute of Physics and Chemistry, Chinese Academy of Sciences, Beijing 100190, China.
   \\$^{5}$Songshan Lake Materials Laboratory, Dongguan, Guangdong 523808, China.
   \\$^{6}$Beijing Academy of Quantum Information Sciences, Beijing 100193, China.
   \\$^{\dag}$These authors contributed equally to this work.
   \\$^{*}$Corresponding author: LZhao@iphy.ac.cn, XJZhou@iphy.ac.cn
}

\date{\today}
\maketitle
\noindent{\bf

High temperature superconductivity in cuprate superconductors is generally considered to be generated from doping the Mott insulators. The fundamental nature of the doped parent compounds as well as the microscopic origin of electron pairing remain critical issues in understanding the emergence of superconductivity. Here, using high-resolution spatially-resolved laser angle-resolved photoemission spectroscopy, we investigate the intrinsic electronic structures of the CuO$_2$ planes in multilayer cuprates $\mathbf{Bi_2 Sr_2 Ca_{n-1} Cu_n O_{2n+4+\delta}}$ (n=$\mathbf{5\sim8}$). The inner CuO$_2$ planes are well shielded from the disorders and provide a rare and ideal platform to probe the intrinsic electronic phase diagram. We observe well-defined Fermi pockets with hole doping levels as low as 0.007, demonstrating an abrupt transition from the parent Mott insulator to a metallic state upon the introduction of an infinitesimal amount of doping. The innermost CuO$_2$ planes ($\mathbf{IP_0}$) display gapless Fermi pockets, while the second innermost planes ($\mathbf{IP_1}$) exhibit anisotropic superconducting gaps up to $\mathbf{\sim33\,\mathbf{meV}}$, indicative of robust electron pairing coexisting with strong antiferromagnetic order. Our findings provide a revised framework for understanding the doping-driven transitions and pairing mechanisms in cuprate superconductors.

}

\vspace{3mm}


\noindent{\bf \large{Introduction}}\\
In nearly forty years of extensive study, a general electronic phase diagram is established for the high temperature cuprate superconductors\cite{dagotto_correlated_1994, damascelli_angle-resolved_2003, lee_doping_2006, keimer_quantum_2015, sobota_angle-resolved_2021}. In this picture, the undoped parent compound is an antiferromagnetic Mott insulator. Upon slight hole doping, the long range antiferromagnetic order is strongly suppressed, but the material stays in the insulating state in the antiferromagnetic region\cite{ando_mobility_2001, kohsaka_growth_2002, ono_evolution_2003, ando_electronic_2004, yu_high-temperature_2019}. Metallic state and superconductivity emerge only after a sufficient amount of doping is introduced and the antiferromagnetic order is fully suppressed. In terms of electronic structure, the strong electron correlation causes a band splitting into the upper Hubbard band (UHB) and lower Hubbard band(LHB) in the parent compound with an intermediate charge transfer band (CTB), forming a charge transfer gap between the charge transfer band and the upper Hubbard band. With a slight hole doping, the in-gap states initially develop around the nodal region but the states are still gapped in the low doping region\cite{shen_fully_2004, vishik_phase_2012, peng_disappearance_2013}. Further doping leads to the formation of Fermi arc around the nodal region and the Fermi arc length increases with increasing doping, eventually forming a large Fermi surface in the overdoped region\cite{ronning_evolution_2003, shen_missing_2004, shen_nodal_2005, yoshida_systematic_2006, hashimoto_doping_2008, tanaka_evolution_2010, peng_disappearance_2013, gao_electronic_2020}.

The recent observation of Fermi pockets in multilayer cuprates has posed serious challenges to the above general pictures\cite{kunisada_observation_2020, kurokawa_unveiling_2023}. In the inner CuO$_2$ planes of the five- and six-layered cuprates Ba$_2$Ca$_{n-1}$Cu$_n$O$_{2n}$(F,O)$_2$\,(n=5 and 6) wihch are far away from disorders, small Fermi pockets are clearly observed centered around ($\pi$/2, $\pi$/2) in the Brillouin zone. These pockets are gapless when their doping levels are smaller than 0.04, indicating the realization of a metallic state in the slightly-doped CuO$_2$ planes. These observations raise a critical question of whether the commonly accepted phase diagram of cuprates is intrinsic or significantly influenced by disorders, particularly regarding whether the slightly-doped CuO$_2$ plane is inherently insulating or metallic in the absence of disorder. A new phase diagram is urgently required to be established that encompasses a wide range of hole doping levels in the disorder-free CuO$_2$ planes. Since no reports of Fermi pocket formation have been made in three-layer ($n=3$)\cite{feng_electronic_2002, ideta_enhanced_2010, luo_electronic_2023} and four-layer ($n=4$)\cite{chen_anomalous_2006} cuprates so far, the multilayer cuprates with $n\geq5$ provide an ideal platform to serve the purpose because the inner CuO$_2$ planes are protected from disorders. However, the multilayer cuprates ($n\geq5$) with single crystals available are scarce and the hole doping range of the inner CuO$_2$ planes is rather limited (<0.05)\cite{kunisada_observation_2020, kurokawa_unveiling_2023}.

In this paper, we report our high-resolution angle-resolved photoemission spectroscopy (ARPES) measurements of the multilayer Bi-based cuprates Bi$_2$Sr$_2$Ca$_{n-1}$Cu$_n$O$_{2n+4+\delta}$ with n=5, 6, 7 and 8. We clearly observed Fermi pockets from the innermost and the second innermost CuO$_2$ planes in all the Bi-based cuprates with different numbers of CuO$_2$ planes (n=$5\sim8$). The hole doping level of the observed Fermi pockets spans a wide range of 0.007 to 0.088. The Fermi pocket from the innermost CuO$_2$ plane(s) is gapless while the Fermi pocket from the second innermost CuO$_2$ plane(s) opens an energy gap which can reach over 30\,meV. These observations provide crucial insights for establishing an intrinsic electronic phase diagram of doping the CuO$_2$ planes in cuprate superconductors.

\begin{figure*}[tbp]
\begin{center}
\includegraphics[width=1\textwidth,angle=0]{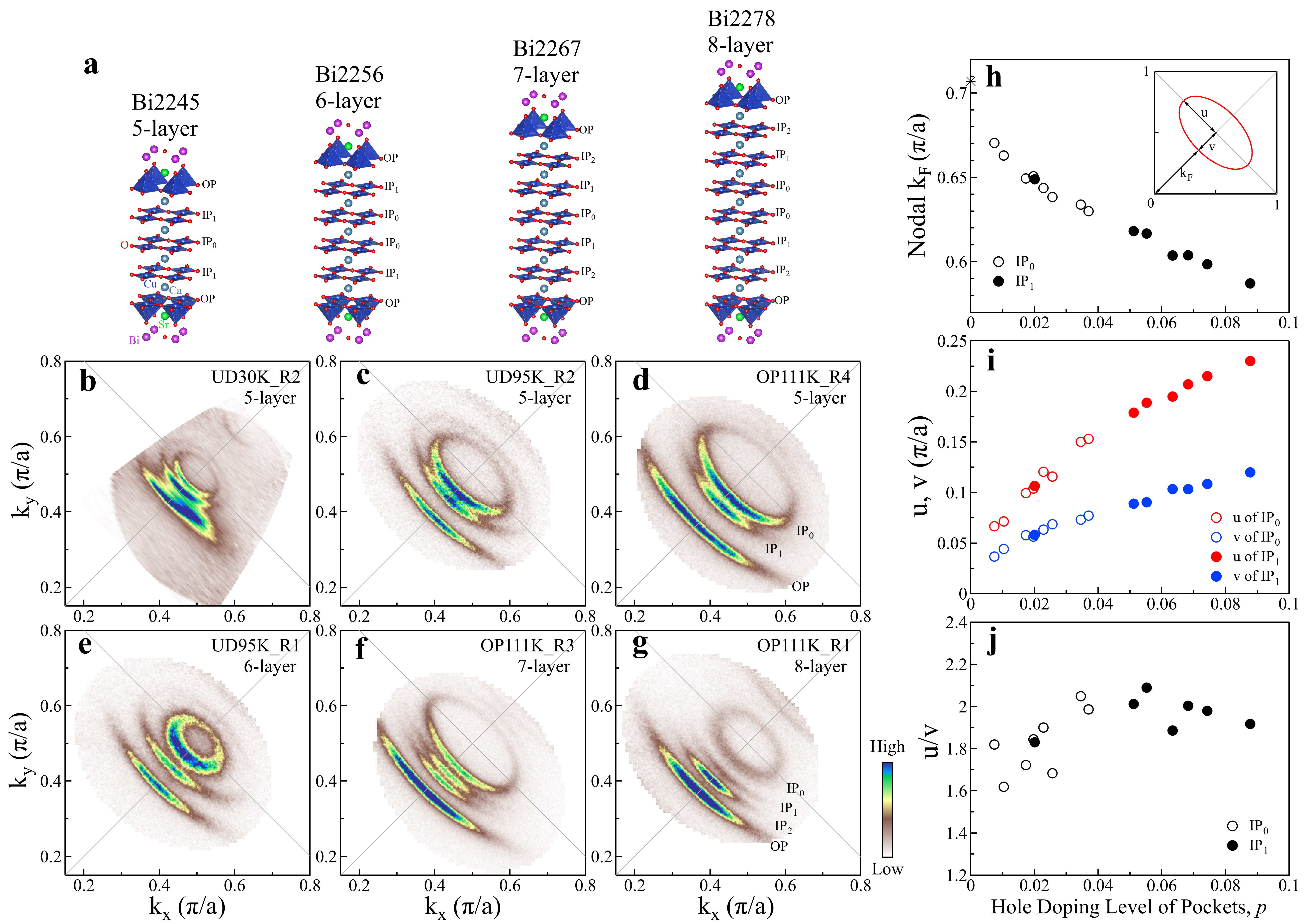}
\end{center}
\caption{{\bf Observation of Fermi pockets in multilayer Bi-based cuprates and their electronic characteristics.}
\textbf{a}, Schematic crystal structures of Bi2245, Bi2256, Bi2267 and Bi2278. The innermost CuO$_2$ plane(s) is labelled as $\mathrm{IP_0}$ while the outermost planes are labelled as OP. The second and third inner planes are labelled as $\mathrm{IP_1}$ and $\mathrm{IP_2}$, respectively.
\textbf{b-g}, Fermi surface mappings measured at 15\,K. Hole-like $\mathrm{IP_0}$ and $\mathrm{IP_1}$ Fermi pockets are clearly observed in region 2 of the UD30K-Bi2223 sample (\textbf{b}, 5-layer), region 2 of the UD95K-Bi2223 sample (\textbf{c}, 5-layer), region 4 of the OP111K-Bi2223 sample (\textbf{d}, 5-layer), region 1 of the UD95K-Bi2223 sample (\textbf{e}, 6-layer), region 3 of the OP111K-Bi2223 sample (\textbf{f}, 7-layer) and region 1 of the OP111K-Bi2223 sample (\textbf{g}, 8-layer). The doping level of each pocket is determined by its enclosed area. 
\textbf{h-j}, Electronic characteristics of the observed $\mathrm{IP_0}$ and $\mathrm{IP_1}$ Fermi pockets. \textbf{h} shows doping dependence of the nodal Fermi momentum $\mathrm{k_F}$ of the Fermi pockets as defined in the upper-right inset. \textbf{i} presents the doping dependence of the major axis length \textit{u} and the minor axis length \textit{v} of the elliptical Fermi pockets. The doping dependence of the axial ratio \textit{u/v} is shown in \textbf{j}.
}
\label{FS}
\end{figure*}

\vspace{3mm}
\noindent{\bf \large{Results}}\\
It has been shown that the Bi$_2$Sr$_2$Ca$_2$Cu$_3$O$_{10+\delta}$ (Bi2223) single crystals may contain intergrowth phases with different numbers of CuO$_2$ planes\cite{wang_correlating_2023}. We successfully measured these intergrowth phases by carrying out spatially-resolved ARPES measurements on the Bi2223 sample surface (see Methods). This achievement is possible thanks to our recent advances in a laser ARPES system, which features a small laser spot and the capability to simultaneously cover two-dimensional momentum space for real-time visualization of the Fermi surface (see Methods and Fig. S1)\cite{zhou_new_2018}. By scanning the sample surface of Bi2223 with different doping levels, we have identified the regions of multilayer Bi-based cuprates with n=5, 6, 7 and 8 (see Fig. \ref{FS}a for their crystal structures) with various doping levels.

Figure \ref{FS}b-g display typical Fermi surface mappings with clear Fermi pockets measured in different regions of various Bi2223 samples with different dopings and T$_c$s. Three distinct Fermi surface sheets are observed in Fig. \ref{FS}b-e, labelled as $\mathrm{IP_0}$, $\mathrm{IP_1}$, and OP, as illustrated in Fig. \ref{FS}d, whereas four Fermi surface sheets are observed in Fig. \ref{FS}f-g, labelled as $\mathrm{IP_0}$, $\mathrm{IP_1}$, $\mathrm{IP_2}$, and OP, as shown in Fig. \ref{FS}g. In all these cases, the $\mathrm{IP_0}$ and $\mathrm{IP_1}$ Fermi surface sheets are clearly Fermi pockets centered around ($\pi$/2,$\pi$/2). As shown from the previous ARPES measurements on multilayer cuprates\cite{kunisada_observation_2020, kurokawa_unveiling_2023, luo_electronic_2023}, the number of observed Fermi surface sheets corresponds mainly to the number of types of the CuO$_2$ planes in the unit cell. Therefore, three Fermi surface sheets (Fig. \ref{FS}b-e) may correspond to five- or six-layer cuprates (Fig. \ref{FS}a) which have three types of CuO$_2$ planes ($\mathrm{IP_0}$, $\mathrm{IP_1}$ and OP) while four Fermi surface sheets (Fig. \ref{FS}f-g) may correspond to seven- or eight-layer cuprates (Fig. \ref{FS}a) which have four types of CuO$_2$ planes ($\mathrm{IP_0}$, $\mathrm{IP_1}$, $\mathrm{IP_2}$ and OP). It is also known from previous measurements that the hole doping level varies among CuO$_2$ planes; it is the highest in the outer planes (OP), gradually decreases in the inner planes ($\mathrm{IP_2}$ and $\mathrm{IP_1}$) and is the lowest in the innermost plane(s) ($\mathrm{IP_0}$)\cite{mukuda_high-tc_2012, kunisada_observation_2020, kurokawa_unveiling_2023, wang_correlating_2023}. This enables the attribution of the observed Fermi surface sheets to the specific CuO$_2$ planes. The $\mathrm{IP_0}$ Fermi pockets in Fig. \ref{FS}b-g originate from the innermost CuO$_2$ planes while the $\mathrm{IP_1}$ Fermi pockets originate from the second innermost CuO$_2$ planes. It has been further shown that the relative size of the $\mathrm{IP_0}$ and $\mathrm{IP_1}$ Fermi pockets can be used to distinguish between an even or odd number of CuO$_2$ planes. When the number of CuO$_2$ planes changes from five to six, it was found that the $\mathrm{IP_0}$ pocket in six-layer cuprate shrinks to half of the $\mathrm{IP_0}$ size in five-layer cuprate because the same hole doping on the innermost plane in five-layer cuprate is divided into the two innermost planes in six-layer cuprate\cite{kunisada_observation_2020, kurokawa_unveiling_2023}. Based on these observations, we can attribute the results in Fig. \ref{FS}b-d to the five-layer cuprates Bi2245 and the result in Fig. \ref{FS}e to the six-layer cuprate Bi2256 because, as seen in Fig. \ref{FS}c and Fig. \ref{FS}e, although OP and $\mathrm{IP_1}$ Fermi surface sheets are similar, the size of the $\mathrm{IP_0}$ Fermi pocket in Fig. \ref{FS}e is nearly half of that in Fig. \ref{FS}c. For the same reason, the result in Fig. \ref{FS}f is attributed to the seven-layer cuprate Bi2267 while the result in Fig. \ref{FS}g is attributed to the eight-layer cuprate Bi2278.

By extensively scanning over many Bi2223 samples, we have successfully measured the electronic structures of multilayer cuprates with n=5,6,7 and 8. Also by measuring Bi2223 samples with different dopings, we can vary the doping levels of the measured multilayer cuprates (see Fig. \ref{FS}b-d). Invariably, the $\mathrm{IP_0}$ and $\mathrm{IP_1}$ Fermi pockets are always observed, as shown in Fig. \ref{FS}b-g and listed in Table S1. The hole doping level of the Fermi pocket is determined from its enclosed area. The smallest $\mathrm{IP_0}$ Fermi pocket with the hole doping level of 0.007 is observed in the five-layer region of the UD30K Bi2223 sample (Fig. \ref{FS}b). The largest $\mathrm{IP_1}$ Fermi pocket with the doping level of 0.088 is observed in the six-layer region of the OP111K Bi2223 sample. Fig. \ref{FS}h-j summarize the electronic characteristics of the observed Fermi pockets. The Fermi momentum decreases monotonically with increasing hole doping level of the Fermi pockets (Fig. \ref{FS}h). All the observed Fermi pockets assume an elliptical shape with their major and minor axes depicted in Fig. \ref{FS}i. Interestingly, although the Fermi pockets size varies in a large range, they keep a similar shape with a slight increase of the ratio between the major axis $u$ and the minor axis $v$ from $u/v$=1.6 to 2.0 over the measured doping range ($p=0.007\sim0.088$)(Fig. \ref{FS}j).

\begin{figure*}[tbp]
\begin{center}
\includegraphics[width=1\textwidth,angle=0]{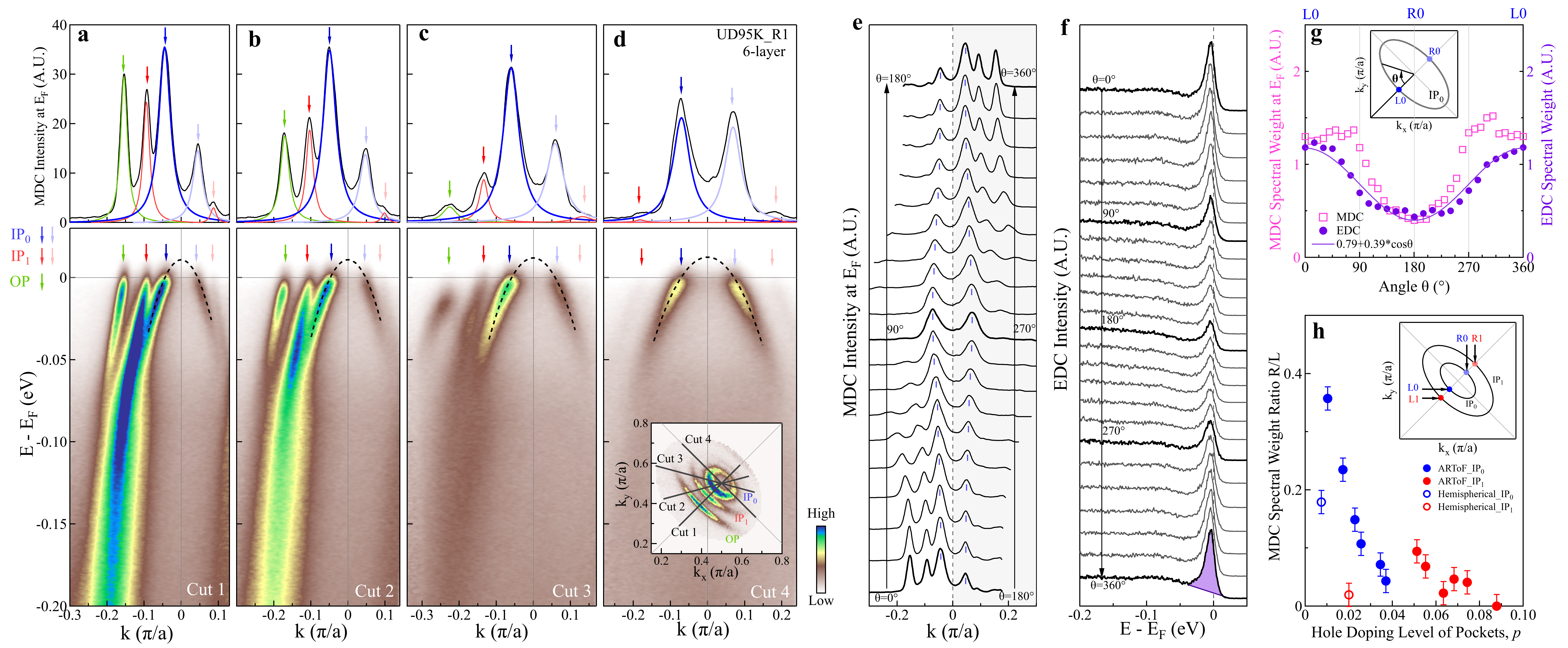}
\end{center}
\caption{{\bf Typical band structures of Fermi pockets and spectral weight distribution along the Fermi pocket measured in region 1 (6-layer) of the UD95K-Bi2223 sample at 15\,K.}
\textbf{a-d}, Band structures measured along different momentum cuts (lower panels) and the corresponding MDCs at the Fermi level (upper panels). The locations of the momentum cuts are indicated by black lines in the inset of \textbf{d}. The observed $\mathrm{IP_0}$, $\mathrm{IP_1}$ and OP bands are marked by blue, red and green arrows, respectively. The $\mathrm{IP_0}$ bands are fitted by parabolic functions (dashed black lines), with the band top located at $\sim$11\,meV. The MDCs are fitted by multiple Lorentzians so each peak area can be obtained. 
\textbf{e}, MDCs at the Fermi level along various momentum cuts intersecting the ($\pi/2, \pi/2$) point. Each cut is characterized by the angle $\theta$, defined in the inset of \textbf{g}.
\textbf{f}, EDCs along the $\mathrm{IP_0}$ Fermi pocket, with the Fermi momentum positions represented by the angle $\theta$ (inset of \textbf{g}).
\textbf{g}, Spectral weight distribution along the $\mathrm{IP_0}$ Fermi pocket, extracted using two methods. The first uses MDC peak areas from \textbf{e}, plotted as pink squares, and the second uses EDC peak areas from \textbf{f}, obtained after background subtraction (as exemplified by the bottom curve), plotted as purple circles. The EDC-derived spectral weight follows an $a + b \cos\theta$ dependence, with $a = 0.79$ and $b = 0.39$ (purple curve). Inset: definition of the Fermi pocket angle $\theta$, with the two minor-axis vertices of the $\mathrm{IP_0}$ pocket labelled as L0 ($\theta = 0^\circ$, $360^\circ$) and R0 ($\theta = 180^\circ$).
\textbf{h}, Spectral weight ratio between the right and left sides of the Fermi pockets for various doping levels. Along the nodal direction, where the energy gap vanishes, the left (L0/L1) and right (R0/R1) vertices of the $\mathrm{IP_0}$ and $\mathrm{IP_1}$ pockets are compared. The spectral weight ratios R0/L0 (blue circles) and R1/L1 (red circles) are derived from the MDC peak intensities at the Fermi level, as shown in Fig. S2. The error bars reflect the uncertainty in determining the spectral weight.
}
\label{Band}
\end{figure*}

Figure \ref{Band} shows typical band structures of the Fermi pockets measured in region 1 (six-layer) of the UD95K-Bi2223 sample at 15\,K. The $\mathrm{IP_0}$, $\mathrm{IP_1}$ and OP bands are clearly observed as marked by the colored arrows in Fig. \ref{Band}a-d. The corresponding momentum distribution curves (MDCs) at the Fermi level are shown in the upper panels of Fig. \ref{Band}a-d. These MDCs are fitted using multiple Lorentzian functions, enabling a quantitative analysis of the spectral weight of the observed bands from the MDC peak area. The $\mathrm{IP_0}$ band exhibits a parabolic dispersion, as highlighted by the dashed black lines, with the band top located at $\sim11\,\mathrm{meV}$ (lower panels of Fig. \ref{Band}a-d). The $\mathrm{IP_1}$ band (left side of the ($\pi$,0)-(0,$\pi$) line) is strong and well-defined near the nodal direction (Cut 1 and Cut 2), but rapidly loses intensity as the momentum cut moves away from the nodal direction (Cut 3) and becomes very weak along the ($\pi$,0)-(0,$\pi$) direction (Cut 4). The OP band (also on the left side of the ($\pi$,0)-(0,$\pi$) line) is relatively strong along the nodal direction (Cut 1) and similarly exhibits a loss of intensity as the momentum cut moves away from the nodal direction (Cut 2 and Cut 3), accompanied by a shift of the band to higher binding energy due to the energy gap opening.

Now we choose the $\mathrm{IP_0}$ Fermi pocket to quantitatively analyze the spectral weight distribution along the entire pocket. To this end, we employ two methods to represent the spectral weight: one is the peak area of  the MDCs at the Fermi level and the other is the peak area of the energy distribution curves (EDCs) along the Fermi pocket. Fig. \ref{Band}e shows MDCs at the Fermi level along different momentum cuts (defined by the Fermi pocket angle $\theta$ in the inset of Fig. \ref{Band}g) centered around ($\pi$/2,$\pi$/2) from $\theta=0^\circ$ to $\theta=180^\circ$ (left side of Fig. \ref{Band}e) and from $\theta=180^\circ$ to $\theta=360^\circ$ (right side of Fig. \ref{Band}e). The MDC peaks corresponding to the $\mathrm{IP_0}$ band are marked by ticks in Fig. \ref{Band}e and their areas are plotted in Fig. \ref{Band}g (pink squares) as a function of the Fermi pocket angle $\theta$. Fig. \ref{Band}f presents EDCs along the $\mathrm{IP_0}$ Fermi pocket at various Fermi pocket angles ($\theta=0$-360$^\circ$). Well-defined quasiparticle peaks are observed along the entire $\mathrm{IP_0}$ pocket. The EDC peak areas, extracted from EDCs in Fig. \ref{Band}f after the background substraction as highlighted in the bottom EDC, are also plotted in Fig. \ref{Band}g (purple circles) as a function of $\theta$. As shown in Fig. \ref{Band}g, the EDC spectral weight decreases from the left minor axis vertex (L0, $\theta=0^\circ$ or 360$^\circ$) to the right minor axis vertex (R0, $\theta=180^\circ$), indicating maximal spectral intensity at L0 and minimal intensity at R0. The EDC spectral weight distribution basically follows the $a + b \cos\theta$ relation with $a=0.79$ and $b=0.39$ (purple line in Fig. \ref{Band}g).

Further analysis of all the observed Fermi pockets reveals that the spectral weight distribution is also strongly dependent on their corresponding doping levels. Fig. \ref{Band}h shows the spectral weight ratio between the right and left minor axis vertices for the $\mathrm{IP_0}$ and $\mathrm{IP_1}$ Fermi pockets (see Fig. S2 for details). The spectral intensity at the right minor axis vertices (R0 and R1 for $\mathrm{IP_0}$ and $\mathrm{IP_1}$, respectively; see inset of Fig. \ref{Band}h) relative to that at the left minor axis vertices (L0 and L1 for $\mathrm{IP_0}$ and $\mathrm{IP_1}$, respectively) decreases rapidly with increasing Fermi pocket size. For the tiny $\mathrm{IP_0}$ pocket with a doping level of $\sim$0.01, the R/L intensity ratio reaches $\sim$36\%, whereas it approaches zero as the doping level increases to $\sim$0.08. We note that our ARPES system equipped with an ARToF electron energy analyzer offers significant advantages in detecting Fermi pockets compared to the hemispherical analyzer, particularly because the spectral intensity on the right side of the pockets is substantially weaker than that on the left. As shown in Fig. \ref{Band}h and Fig. S2, for Fermi pockets with similar doping levels, the spectral weight at the right minor axis vertices relative to the left is consistently higher when measured with the ARToF-ARPES system (filled circles) than with the system with the hemispherical analyzer (open circles). This is attributed to the higher signal-to-noise ratio and the absence of nonlinear effects in the ARToF analyzer, in contrast to the hemispherical analyzer. Therefore, the spectral weight distribution acquired using the ARToF analyzer is considered more intrinsic than that obtained using the hemispherical analyzer.

\begin{figure*}[tbp]
\begin{center}
\includegraphics[width=1\textwidth,angle=0]{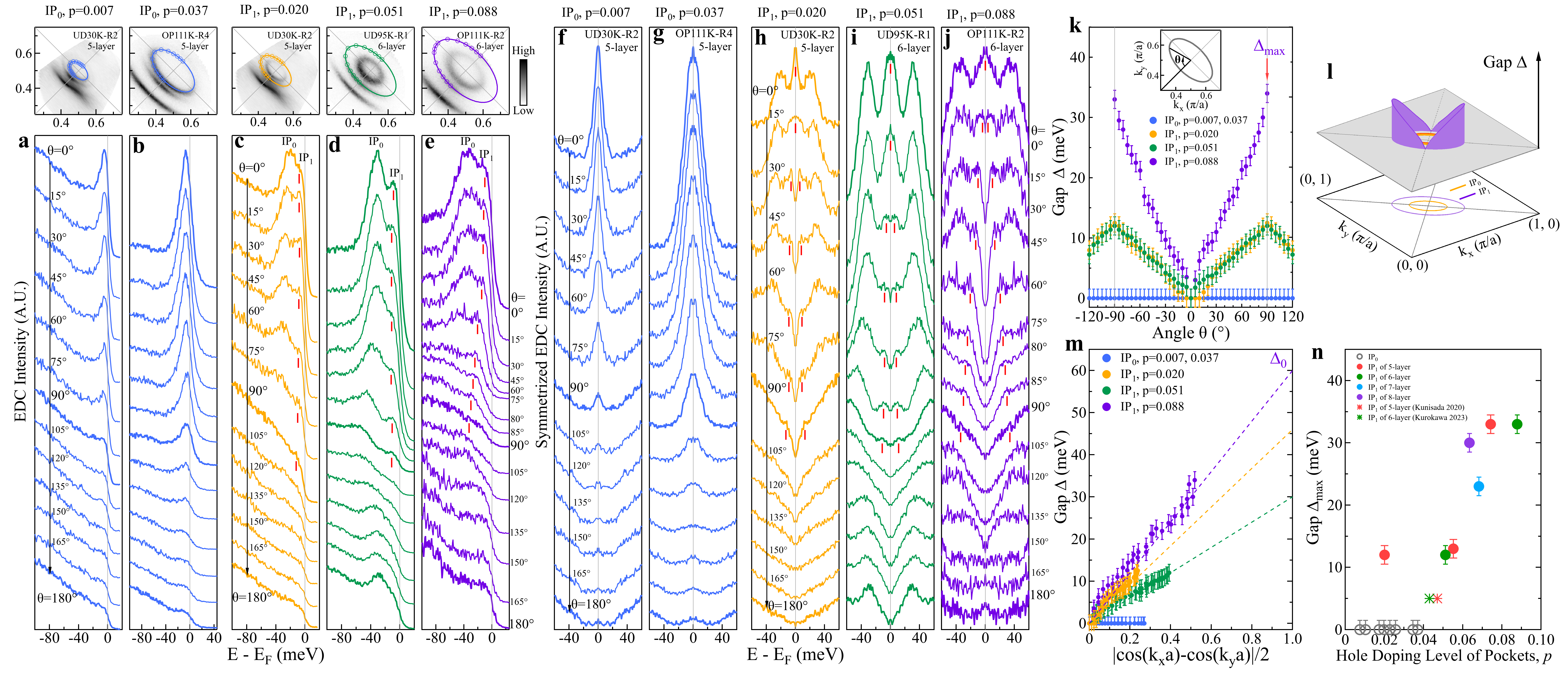}
\end{center}
\caption{{\bf Momentum-dependent photoemission spectra and energy gaps along the Fermi pockets measured at 15\,K.}
\textbf{a,b}, EDCs measured along the $\mathrm{IP_0}$ Fermi pockets in region 2 of the UD30K-Bi2223 sample (\textbf{a}, 5-layer) and region 4 of the OP111K-Bi2223 sample (\textbf{b}, 5-layer). The Fermi momenta are indicated by open circles in the upper panels and represented by the angle $\theta$, as defined in the inset of \textbf{k}.
\textbf{c-e}, EDCs measured along the $\mathrm{IP_1}$ Fermi pockets in region 2 of the UD30K-Bi2223 sample (\textbf{c}, 5-layer), region 1 of the UD95K-Bi2223 sample (\textbf{d}, 6-layer) and region 2 of the OP111K-Bi2223 sample (\textbf{e}, 6-layer). Fermi momentum positions are marked by open circles and represented by the angle $\theta$.
\textbf{f-j}, Symmetrized EDCs from \textbf{a-e}.
\textbf{k}, Momentum-dependent energy gaps along the $\mathrm{IP_0}$ and $\mathrm{IP_1}$ Fermi pockets plotted as a function of the angle $\theta$. Gap values are extracted from the symmetrized EDCs in \textbf{f-j}.
\textbf{l}, Schematic three-dimensional representation of energy gaps along the $\mathrm{IP_0}$ and $\mathrm{IP_1}$ Fermi pockets (top layer), with the corresponding Fermi pocket contours displayed in the bottom layer.
\textbf{m}, Energy gaps from \textbf{k} replotted as a function of $|\cos(k_x a) - \cos(k_y a)|/2$.
\textbf{n}, Maximum energy gaps for all observed Fermi pockets. Data from previously reported Fermi pockets are included for comparison (asterisks)\cite{kunisada_observation_2020,kurokawa_unveiling_2023}. The error bars reflect the uncertainty in determining the gap size.
}
\label{EDC}
\end{figure*}

We now turn to the investigation of gap opening along the Fermi pockets. Fig. \ref{EDC}a-e show EDCs measured along two representative $\mathrm{IP_0}$ Fermi pockets (lower panels of Fig. \ref{EDC}a,b) and three representative $\mathrm{IP_1}$ Fermi pockets (lower panels of Fig. \ref{EDC}c-e). The corresponding symmetrized EDCs are presented in Fig. \ref{EDC}f-j. Well-defined peaks are observed in EDCs measured on the left side of the $\mathrm{IP_0}$ pockets ($\theta=0^\circ$–$90^\circ$ in Fig. \ref{EDC}a,b), whereas the peaks are significantly weaker on the right side ($\theta=90^\circ$–$180^\circ$) due to the low spectral weight on the right side (see Fig. \ref{Band}h). The EDCs along the $\mathrm{IP_1}$ pockets (lower panels of Fig. \ref{EDC}c-e) exhibit two features: the component near the Fermi level (marked by ticks) corresponds to the $\mathrm{IP_1}$ band, while the higher binding energy component originates from the overlapping $\mathrm{IP_0}$ band. Overall, the EDC peaks of the $\mathrm{IP_1}$ band in Fig. \ref{EDC}c-e appear broader and weaker in intensity compared to those of the $\mathrm{IP_0}$ band in Fig. \ref{EDC}a,b.

The energy gap along the Fermi pocket is extracted from the symmetrized EDCs (Fig. \ref{EDC}f-j). For both $\mathrm{IP_0}$ pockets with doping levels of 0.007 (Fig. \ref{EDC}f) and 0.037 (Fig. \ref{EDC}g), all symmetrized EDCs show a peak at the Fermi level, indicating the absence of gap opening along these pockets. For the $\mathrm{IP_1}$ pockets with doping levels of 0.020 (Fig. \ref{EDC}h), 0.051 (Fig. \ref{EDC}i) and 0.088 (Fig. \ref{EDC}j), the symmetrized EDCs along the nodal direction ($\theta=0^\circ$) also show a peak at the Fermi level indicating zero gap. However, with increasing Fermi pocket angle $\theta$ from 0$^\circ$ to 90$^\circ$, the symmetrized EDCs progressively develop a dip at the Fermi level and the quasiparticle peaks shift away from $E_F$, reflecting a growing energy gap (Fig. \ref{EDC}h-j). The gap amplitudes obtained from Fig. \ref{EDC}f-j are summarized as a function of $\theta$ in Fig. \ref{EDC}k. The $\mathrm{IP_1}$ pockets exhibit a highly anisotropic gap, with a minimum (zero) at the minor axis vertices ($\theta=0^\circ$) and a maximum at the major axis vertices ($\theta=\pm90^\circ$), whereas no gap is detected on any of the observed $\mathrm{IP_0}$ pockets. This gap distribution is schematically illustrated in the three-dimensional rendering in Fig. \ref{EDC}l. To check whether the momentum dependence of the energy gap follows a d-wave form, we plot the gap values along the $\mathrm{IP_1}$ Fermi pockets as a function of |cos(k$_x$a)-cos(k$_y$a)|/2 in Fig. \ref{EDC}m. The energy gaps of the three $\mathrm{IP_1}$ pockets (with doping levels 0.020, 0.051, and 0.088) all conform to a $d$-wave form, with extrapolated maximum gaps $\Delta_0$ of 46, 30, and 60\,meV, respectively.

Figure \ref{EDC}n summarizes the maximum energy gap for all observed $\mathrm{IP_0}$ and $\mathrm{IP_1}$ Fermi pockets. Our observed $\mathrm{IP_0}$ pockets cover a doping range of 0.007-0.037 and there is no gap opening on the $\mathrm{IP_0}$ pockets. This is consistent with the previous observation of zero gap on the $\mathrm{IP_0}$ pockets in the five-layer and six-layer cuprates\cite{kunisada_observation_2020, kurokawa_unveiling_2023}. In contrast, the $\mathrm{IP_1}$ pockets observed here span 0.020–0.088 in doping and all exhibit a finite energy gap. Notably, gap opening is already evident even on the smallest $\mathrm{IP_1}$ pocket at doping 0.020 (Fig. \ref{EDC}n and Fig. S3). This is different from the earlier report that there is no gap opening on the $\mathrm{IP_1}$ pocket when the doping level is less than 0.04\cite{kurokawa_unveiling_2023}. Our measurements further reveal that the maximum gap on the $\mathrm{IP_1}$ pockets reaches up to 33\,meV (Fig. \ref{EDC}n). This is substantially larger than $5$\,meV gaps previously reported on the $\mathrm{IP_1}$ pockets in the five-layer and six-layer cuprates\cite{kunisada_observation_2020, kurokawa_unveiling_2023}. These contrasting behaviors between $\mathrm{IP_0}$ and $\mathrm{IP_1}$ pockets, as well as among $\mathrm{IP_1}$ pockets in different systems, indicate that gap opening and magnitude are not governed solely by the doping level. Many additional factors need to be considered including the antiferromagnetic order and interlayer interactions.

It has been found that the antiferromagnetic order is strongly enhanced in multilayer cuprates; the strength of the antiferromagnetic order increases with the number of CuO$_2$ layers in a structural unit (n)\cite{mukuda_high-tc_2012}. The formation of the Fermi pockets in our Bi-based multilayer cuprates (n=5$\sim$8) arises from the coexistence of pronounced electron correlations and strong antiferromagnetic order in the lightly doped regime. In principle, the electronic structure of the system can be described by considering the in-plane electron dynamics of each CuO$_2$ layer and the interlayer coupling. This interlayer coupling is highly sensitive to the doping levels of adjacent layers and is known to increase with doping\cite{mori_fermi_2006}. Since the inner layers in our multilayer cuprates are underdoped or heavily underdoped, the interlayer hopping between the inner layers is rather weak. It has been shown that, when interlayer hopping is small, the Fermi surface topology and associated band structures of the $\mathrm{IP_0}$ and $\mathrm{IP_1}$ Fermi pockets are primarily governed by the innermost and second innermost CuO$_2$ layers, respectively, with minimal influence from interlayer coupling\cite{kurokawa_unveiling_2023}. Therefore, to understand the formation of the $\mathrm{IP_0}$ and $\mathrm{IP_1}$ Fermi pockets, it suffices to treat the electronic structure of each CuO$_2$ layer independently.

\begin{figure*}[tbp]
\begin{center}
\includegraphics[width=1\textwidth,angle=0]{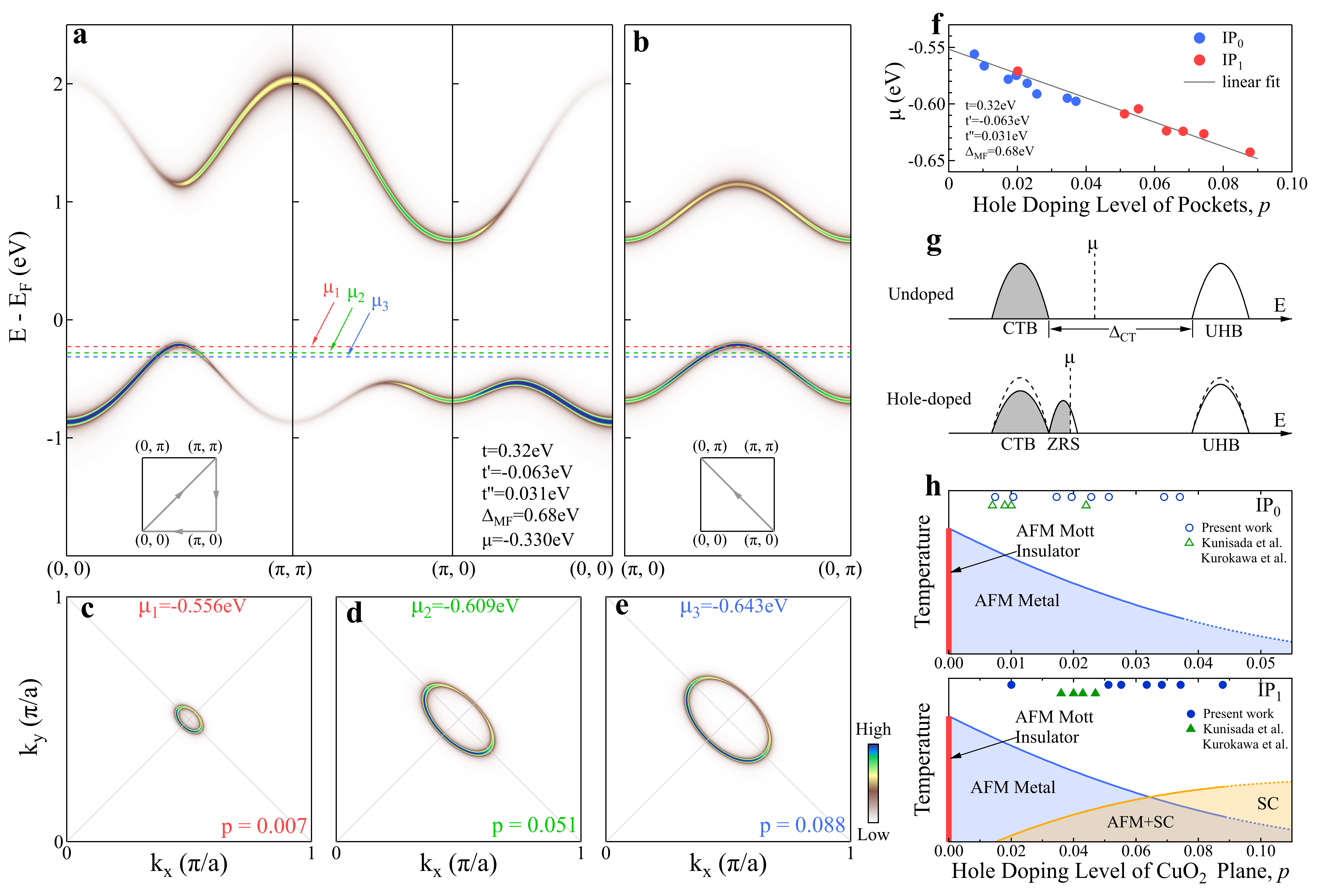}
\end{center}
\caption{{\bf Formation of the Fermi pockets in terms of the mean-field t-U model and intrinsic doping evolution of the CuO$_2$ plane in the antiferromagnetic Mott insulating state.}
\textbf{a,b}, Simulated band structures along high-symmetry directions (0,0)–($\pi$,$\pi$)–($\pi$,0)–(0,0) (\textbf{a}) and ($\pi$,0)–(0,$\pi$) (\textbf{b}) based on the mean-field $t$–$U$ model using $t = 0.32$\,eV, $t' = -0.063$\,eV, $t'' = 0.031$\,eV, and $\Delta_{\mathrm{MF}} = 0.68$\,eV.
\textbf{c-e}, Simulated Fermi pockets derived from the band structures in \textbf{a,b} by setting the chemical potential to $\mu_1 = -0.556$\,eV (\textbf{c}), $\mu_2 = -0.609$\,eV (\textbf{d}), and $\mu_3 = -0.643$\,eV (\textbf{e}), as indicated by dashed lines in \textbf{a,b}. These simulated pockets correspond to doping levels of 0.007 (\textbf{c}), 0.051 (\textbf{d}), and 0.088 (\textbf{e}), respectively.
\textbf{f}, Doping dependence of the extracted chemical potential $\mu$, obtained by fitting the experimentally observed Fermi pockets and band structures using the mean-field $t$–$U$ model. The hopping integrals $t$, $t'$, $t''$, and $\Delta_{\mathrm{MF}}$ are fixed as listed in the figure.
\textbf{g}, Schematic illustration of the electronic structure for undoped (upper panel) and lightly hole-doped (lower panel) CuO$_2$ planes. In the undoped case, the charge transfer band (CTB) and upper Hubbard band (UHB) are separated by a charge transfer gap $\Delta_{\mathrm{CT}}$. With slight hole doping, the Zhang-Rice singlet (ZRS) band emerges and the chemical potential shifts to the top of the ZRS band.
\textbf{h}, Schematic phase diagrams showing the intrinsic doping evolution of the $\mathrm{IP_0}$ (upper panel) and $\mathrm{IP_1}$ (lower panel) CuO$_2$ planes. The red bar denotes the antiferromagnetic Mott insulating phase at zero doping. The blue and yellow shaded regions indicate the antiferromagnetic metallic and superconducting phases, respectively. Doping levels of the observed $\mathrm{IP_0}$ and $\mathrm{IP_1}$ Fermi pockets are marked by blue open and filled circles, respectively. Previously reported Fermi pocket data are shown as green open triangles ($\mathrm{IP_0}$) and green filled triangles ($\mathrm{IP_1}$)\cite{kunisada_observation_2020, kurokawa_unveiling_2023}.
}
\label{Simu}
\end{figure*}

We find that the measured $\mathrm{IP_0}$ and $\mathrm{IP_1}$ Fermi pockets and the associated band structures can be well described by the mean-field $t$–$U$ model\cite{schrieffer_dynamic_1989, chubukov_renormalized_1992,bulut_electronic_1994, xiang_quasiparticle_1996, kusko_fermi_2002}. This model involves in-plane hopping integrals $t$, $t'$, and $t''$, as well as a mean-field energy gap $\mathrm{\Delta_{MF}}$, as detailed in the Supplementary Materials. We fitted the Fermi surface and associated band structures along three representative momentum cuts ($\theta = 0^\circ$, $45^\circ$, and $90^\circ$) of all observed $\mathrm{IP_0}$ and $\mathrm{IP_1}$ Fermi pockets using the same set of parameters \{$t$, $t'$, $t''$, $\mathrm{\Delta_{MF}}$\} = \{0.32\,eV, –0.063\,eV, 0.031\,eV, 0.68\,eV\}, while varying only the chemical potential $\mu$. The fitted values of $\mu$ exhibit an approximately linear dependence on the doping level of the Fermi pockets (Fig. \ref{Simu}f). This result suggests that a single parameter set \{$t$, $t'$, $t''$, $\mathrm{\Delta_{MF}}$\}, together with different values of $\mu$, is sufficient to capture the electronic structure of all observed $\mathrm{IP_0}$ and $\mathrm{IP_1}$ pockets. Fig. \ref{Simu}a,b presents the simulated band structures along high-symmetry directions based on these parameters. By adjusting only the chemical potential, the band structures in Fig. \ref{Simu}a,b reproduce the formation and characteristics of all observed $\mathrm{IP_0}$ and $\mathrm{IP_1}$ Fermi pockets. As exemplified in Fig. S4, the simulated Fermi pockets and band structures obtained from these fitted parameters are in good agreement with the experimental data. The spectral weight distribution along the simulated pocket (Fig. S5) also closely resembles that of the experimental measurements (Fig. \ref{Band}g). As shown in Fig. \ref{Simu}a,b, by setting the chemical potential to different values ($\mu_1 = -0.556$\,eV, $\mu_2 = -0.609$\,eV, $\mu_3 = -0.643$\,eV), Fermi pockets of various sizes can be obtained, as displayed in Fig. \ref{Simu}c-e, corresponding to hole doping levels of 0.007, 0.051, and 0.088, respectively. The size and shape of these simulated pockets show excellent agreement with those observed in the experimental data (Fig. \ref{FS}b-g).

\vspace{3mm}
\noindent{\bf \large{Discussion}}\\
Our present study offers critical insights into the intrinsic nature of the lightly doped CuO$_2$ plane. The fundamental question is whether the CuO$_2$ plane remains insulating or becomes metallic upon slight hole doping. In the general electronic phase diagram established so far, in-plane insulating behaviors have been reported in various slightly hole-doped cuprates with doping levels $p < 0.03$\cite{ando_mobility_2001, kohsaka_growth_2002, ono_evolution_2003, ando_electronic_2004, keimer_quantum_2015, yu_high-temperature_2019}. ARPES measurements detect little spectral weight at the Fermi level in such lightly doped systems\cite{shen_missing_2004, peng_disappearance_2013, gao_electronic_2020}. These results point to the insulating nature of the CuO$_2$ plane with a slight hole doping. In contrast, our results demonstrate that well-defined Fermi pockets ($\mathrm{IP_0}$) form with sharp EDC peaks even at doping levels as low as 0.01 (Fig. \ref{FS}e and Fig. \ref{Band}f) or 0.007 (Fig. \ref{FS}b and Fig. \ref{EDC}a). This clearly indicates a metallic ground state emerges in the CuO$_2$ plane with minimal hole doping. This discrepancy between previous reports and our findings can be attributed to disorder effects. In conventional single-layer and bilayer cuprates, the CuO$_2$ planes lie adjacent to charge reservoir layers, and the doped holes tend to be localized by disorder from nearby layers\cite{mcelroy_atomic-scale_2005, ye_visualizing_2023}. In our multilayer cuprates ($n=5$–8), the innermost CuO$_2$ plane ($\mathrm{IP_0}$) and second innermost plane ($\mathrm{IP_1}$) are well separated from the charge reservoir layers and thus effectively shielded from disorder. In this disorder-free environment, doped holes are not pinned and can propagate freely. Therefore, the emergence of a metallic state in multilayer cuprates reflects the intrinsic nature of the lightly doped CuO$_2$ plane. Observations of insulating behavior in single-layer and bilayer cuprates must be re-evaluated in light of strong disorder effects. The undoped CuO$_2$ plane is a well-established Mott insulator with an energy gap of $\sim$1\,eV\cite{uchida_optical_1991, ye_visualizing_2013}. Our data reveal that even infinitesimal hole doping triggers the formation of a Fermi pocket, indicating a metallic transition. This suggests an abrupt transformation from a large-gap insulating state to a metallic state upon doping.

Our findings establish a new intrinsic framework for understanding the doping evolution of electronic structures in disorder-free CuO$_2$ planes. In previous studies on single-layer and bilayer cuprates, a new in-gap state emerges near the Fermi level upon slight hole doping of the parent compound. With further doping, this state gradually intensifies and becomes increasingly coherent, eventually evolving into a sharp quasiparticle peak once the doping level surpasses a critical threshold, $\sim$0.07 for Bi2201\cite{peng_disappearance_2013} and Bi2212\cite{gao_electronic_2020}, and $\sim$0.10 for $\mathrm{Ca_{2-x}Na_xCuO_2Cl_2}$\cite{shen_missing_2004}. In contrast, our results reveal that even at an ultralow doping level of 0.007, the chemical potential shifts downward immediately, and a well-defined quasiparticle band with strong coherence—labelled as the Zhang-Rice singlet (ZRS) state—emerges directly at the Fermi level (Fig. \ref{Simu}g). The doping evolution of the Fermi pockets and the associated band structures in the doping range of 0.007-0.088 can be understood by maintaining the same band structures (Fig. \ref{Simu}a,b) but only shifting the chemical potential (Fig. \ref{Simu}f). This observation underscores a qualitatively distinct doping evolution in the electronic structure of an intrinsically doped CuO$_2$ plane.

Furthermore, our findings provide critical insights into the mechanism of electron pairing in cuprate superconductors. It is generally believed that superconductivity emerges only when the long-range antiferromagnetic order is suppressed to residual spin fluctuations\cite{keimer_quantum_2015}. Spin fluctuations have been widely considered a leading candidate for mediating electron pairing in cuprate superconductors\cite{scalapino_d-wave_1986, monthoux_toward_1991, scalapino_case_1995, moriya_spin_2000, abanov_gap_2008, dahm_strength_2009}. However, our proposed intrinsic phase diagram demonstrates that electron pairing can occur across a broad doping range (0.02–0.088), where strong antiferromagnetic order still persists in the multilayer cuprates (Fig. \ref{EDC}n and Fig. \ref{Simu}h). Remarkably, the electron pairing can occur in the $\mathrm{IP_1}$ CuO$_2$ plane with a doping level as low as 0.02 and the energy gaps as large as $\sim$33\,meV can open along the $\mathrm{IP_1}$ pockets at low doping levels where the antiferromagnetic order remains strong (Fig. \ref{EDC}n). These results indicate that electron pairing can coexist with, or even be facilitated by, strong antiferromagnetic correlations. This observation aligns with earlier proposals suggesting that the dominant interaction that gives rise to the electron pairing is the same superexchange spin coupling which causes the undoped cuprates to be antiferromagnetic Mott insulators\cite{anderson_resonating_1987, anderson_last_2016}. Nonetheless, this framework does not fully account for why no energy gap opens on the $\mathrm{IP_0}$ Fermi pockets, even under similarly strong antiferromagnetic conditions and low doping levels (Fig. \ref{EDC}n). Nor does it explain why the gap on the $\mathrm{IP_1}$ pockets decreases as the doping level approaches 0.02 (Fig. \ref{EDC}n). As shown in Fig. \ref{Simu}h, the $\mathrm{IP_0}$ (upper panel) and $\mathrm{IP_1}$ (lower panel) Fermi pockets exhibit distinct doping evolutions. This suggests that electron pairing and superconductivity in the CuO$_2$ planes are not governed solely by the doping level; additional factors such as interlayer coupling, spin interaction strength, and local structural environments must also be considered.

In conclusion, a series of Fermi pockets and their doping- and layer-dependent evolution in multilayer Bi-based cuprates (\(\mathrm{Bi_2Sr_2Ca_{n-1}Cu_nO_{2n+4+\delta}}\), \(n=5\sim8\)) were observed using high-resolution spatially-resolved laser ARPES. These multilayer cuprates offer a rare and invaluable platform for probing the intrinsic electronic phase diagram and superconductivity mechanism. Fermi pockets of varying sizes were detected, with the hole doping level reaching as low as 0.007. This suggests that the undoped insulating CuO\(_2\) plane transitions into a metallic state with the introduction of even a tiny amount of holes. The formation of these Fermi pockets can be attributed to the strong electron correlation and the associated antiferromagnetic order in the disorder-free CuO\(_2\) planes. The innermost CuO\(_2\) planes (\(\mathrm{IP_0}\)) exhibit gapless Fermi pockets, whereas the second innermost planes (\(\mathrm{IP_1}\)) display anisotropic energy gaps as large as 33\,meV. These observations suggest that electron pairing can occur at very low doping levels in the presence of strong antiferromagnetic order. Our findings offer crucial insights into the nature of doped Mott insulators and the pairing mechanism in cuprate superconductors.

\newpage
\noindent{\bf \large{Methods}}\\
\noindent{\bf Sample Preparation}\\
\hspace*{4mm} Single crystals of $\mathrm{Bi_{2}Sr_{2}Ca_{2}Cu_{3}O_{10+\delta}}$ (Bi2223) were grown by the traveling-solvent floating-zone method\cite{lin_growth_2002, liang_single_2002}. The scanning transmission electron microscopy (STEM) analyses reveal that these Bi2223 samples contain intergrowth phases with different numbers of CuO$_2$ planes ($n = 1\sim9$)\cite{wang_correlating_2023}. The as-grown Bi2223 single crystals are underdoped with a $T_c$ at 95\,K (named as UD95K-Bi2223 in the paper). 
These Bi2223 single crystals were post-annealed in a flowing oxygen atmosphere to obtain optimally-doped samples with a $T_c$ at 111\,K (named as OP111K-Bi2223 in the paper). The as-grown Bi2223 single crystals were also annealed in vacuum at 450\,$^\circ C$ for 3\,days to obtain heavily underdoped samples with a $T_c$ at 30\,K (named as UD30K-Bi2223 in the paper). The superconducting transition temperatures ($T_c$) were determined via magnetic susceptibility measurements. As the doping level of the host Bi2223 crystal is tuned, the doping levels of the intergrowth phases are expected to change accordingly.

\vspace{2mm}
\noindent{\bf ARPES Measurements}\\
High-resolution ARPES measurements were carried out using our lab-based ARPES system equipped with a 6.994\,eV vacuum ultraviolet (VUV) laser and an angle-resolved time-of-flight (ARToF) electron energy analyzer, as well as another laser-based ARPES setup utilizing a 6.994\,eV VUV laser and a DA30L hemispherical electron energy analyzer (Scienta-Omicron)\cite{liu_development_2008, zhou_new_2018}. The ARToF ARPES system enables simultaneous acquisition of photoelectrons across two-dimensional momentum space ($k_x$, $k_y$), allowing real-time visualization of the Fermi surface, which is critical for identifying multilayer phases in Bi2223 samples. The energy resolution was set at $\sim$1\,meV, and the angular resolution was approximately 0.3\,degrees, corresponding to a momentum resolution of 0.004\,Å$^{-1}$. The laser spot size used in both ARPES systems was set at $\sim$10\,$\mu$m. All Bi2223 samples were cleaved \textit{in situ} at low temperature and measured in ultrahigh vacuum with a base pressure better than $5\times10^{-11}$ Torr. The Fermi level is referenced by measuring a clean polycrystalline gold which is good electrically connected to the sample, and cross-referenced using nodal electronic states which remain gapless in the superconducting state.

\vspace{2mm}
\noindent{\bf Spatially-Resolved ARPES Measurements}\\
It has been shown that Bi2223 single crystal samples contain intergrowth phases with varying numbers of CuO$_2$ planes ($n = 1\sim9$)\cite{wang_correlating_2023}. To distinguish these phases within the Bi2223 samples, we performed spatially-resolved ARPES measurements across the sample surface. Taking advantage of the two-dimensional momentum coverage of the ARToF ARPES system, we conducted a full spatial scan by acquiring Fermi surface mappings at each point, as illustrated in Supplementary Fig. S1a. Different regions of the surface exhibited distinct Fermi surface topologies. In some regions, one Fermi surface sheet (Fig. S1e) or two Fermi surface sheets (Fig. S1f) can be observed. In some other regions, well-defined Fermi pockets were clearly visible (Fig. S1g,h). The assignment of each Fermi surface pattern to a specific intergrowth phase was made by analyzing the number of Fermi surface sheets, their momentum-space distribution, and corresponding doping levels, as detailed in the main text. This study primarily focuses on the regions exhibiting well-resolved Fermi pockets. The typical size of these regions is $\sim$50\,$\mu$m in the sample surface. We successfully observed Fermi pockets in two regions of the UD30K Bi2223 sample (denoted as UD30K-R1 and UD30K-R2), two regions of the UD95K Bi2223 sample (UD95K-R1 and UD95K-R2), and four regions of the OP111K Bi2223 sample (OP111K-R1, OP111K-R2, OP111K-R3, and OP111K-R4), as listed in Supplementary Table S1. The measurement on region 2 of the UD30K-Bi2223 sample was carried out by using the ARPES system with the hemispherical electron energy analyzer (Fig. 1b) while all the other measurements were carried out by using the ARPES system with the ARToF electron energy analyzer.


\vspace{3mm}



\vspace{8mm}
\noindent{\bf \large{References}}\\

\vspace{3mm}

\noindent {\bf \large{Acknowledgements}}\\
 This work is supported by the National Natural Science Foundation of China (Grant Nos. 12488201 by X.J.Z., 12374066 by L.Z. and 12374154 by X.T.L.), the National Key Research and Development Program of China (Grant Nos. 2021YFA1401800 by X.J.Z., 2022YFA1604200 by L.Z., 2022YFA1403900 by G.D.L. and 2023YFA1406000 by X.T.L.), the Strategic Priority Research Program (B) of the Chinese Academy of Sciences (Grant No. XDB25000000 by X.J.Z.), Innovation Program for Quantum Science and Technology (Grant No. 2021ZD0301800 by X.J.Z.), the Youth Innovation Promotion Association of CAS (Grant No. Y2021006 by L.Z.) and the Synergetic Extreme Condition User Facility (SECUF).

\vspace{3mm}
\noindent {\bf \large{Author Contributions}}\\
 X.J.Z., L.Z., H.C. and J.M.S. proposed and designed the research. H.C. and J.M.S. carried out the ARPES experiments. C.T.L grew the Bi2223 single crystals. H.C. and J.M.S. contributed in sample preparation. X.Y.L., Y.W.C., C.H.Y, Y.J.S., J.X.Z., T.M.M, B.L., W.P.Z, N.C., X.L.R., S.J.Z, Z.M.W, F.F.Z., F.Y., Q.J.P., Z.Y.X., G.D.L., X.T.L., H.Q.M., L.Z. and X.J.Z. contributed to the development and maintenance of the ARPES systems and related software development. Y.H.L., H.C., J.M.S., X.Y.L. and T.X. contributed to theoretical analysis and discussions. H.C., J.M.S., L.Z. and X.J.Z. analyzed the data. X.J.Z., L.Z., H.C. and J.M.S. wrote the paper. All authors participated in discussions and comments on the paper.


\newpage

\clearpage

\clearpage

\clearpage

\end{document}